\documentclass[12pt]{article}
\usepackage{epsfig,amsmath,amssymb}

\def\beq{\begin{equation}}
\def\eeq#1{\label{#1}\end{equation}}
\def\eeqn{\end{equation}}

\def\beqa{\begin{eqnarray}}
\def\eeqa#1{\label{#1}\end{eqnarray}}
\def\eeqan{\end{eqnarray}}

\let\bar=\overbar

\def\tr{{\mbox{\rm tr}}}

\def\Dslash{\not{\hbox{\kern-4pt $D$}}}
\def\dslash{\not{\hbox{\kern-2pt $\del$}}}

\def\msb{{\bar{\ssstyle M \kern -1pt S}}}

\def\Title#1{\begin{center} {\Large {\bf #1} } \end{center}}

\begin{document}

\Title{Large-$N_c$ universality of phases in QCD and QCD-like theories}

\bigskip\bigskip

%+\addtocontents{toc}{{\it D. Reggiano}}
%+\label{ReggianoStart}

\begin{raggedright}  

{\it Masanori  Hanada\\
Department of Physics \\
University of Washington \\
 Seattle, WA 98195-1560, USA}
\bigskip\bigskip
\end{raggedright}

\section{Introduction}

QCD with a finite baryon chemical potential, despite its importance, is not well understood because the standard 
lattice QCD simulation is not applicable due to the sign problem.  
Although sign-free QCD-like theories have been studied intensively, 
relation to QCD with a finite baryon chemical potential was not clear until recently \cite{Cherman:2010jj,Hanada:2011ju}. 
In this talk we explain the large-$N_c$ equivalences between QCD and various QCD-like theories,  
which lead us to a unified viewpoint for QCD with baryon and isospin chemical potentials, $SO(2N_c)$ and $Sp(2N_c)$ gauge theories. 
In particular  two-flavor QCD with the baryon chemical potential is equivalent to  its phase quenched version 
in a certain parameter region, 
which is relevant for heavy ion collision experiments.  
%Furthermore, the leading correction to the large-$N_c$ limit is not sensitive to the sign. 
%From this point of view, simulation results of the phase-quenched theories \cite{Kogut:2007mz,deForcrand:2007uz} 
%suggests the non-existence of the QCD critical point 

\section{Basic Idea}

Consider QCD at a finite baryon chemical potential, 
\begin{eqnarray}
{\cal L} = \frac{1}{4 g^{2} } \tr ({F}_{\mu \nu})^2
+ 
\sum_{f=1}^{N_f} 
\bar{\psi}_{f }\left( \gamma^{\mu} {D}_{\mu} + m_f + \mu \gamma^4 \right)\psi_{f }, 
\end{eqnarray}
where the gauge group is $SU(3)$, $N_f$ is the number of flavors, $\psi_f$ are quarks of mass $m_f$ in the fundamental representation, 
and $\mu$ is the quark chemical potential which is related to the baryon chemical potential $\mu_B$ 
as $\mu_B = 3\mu$.  
This system suffers from the {\it fermion sign problem} -- 
the fermion determinant $\prod_{f=1}^{N_f}\det \left( \gamma^{\mu} {D}_{\mu} + m_f + \mu \gamma^4 \right)$ 
becomes complex, rendering importance sampling impossible in practice. 

In order to circumvent this difficulty, gauge theories which do not suffer from the sign problem at finite density have been studied. 
Consider QCD and QCD-like theories\footnote{
In this paper we call $SU(N_c)$ Yang-Mills with $N_f$ fundamental fermions `QCD'. 
$SU(N_c)$ Yang-Mills with fermions in other representations and $SO(2N_c)$/$Sp(2N_c)$ theories 
are referred to as `QCD-like theories'.  
} of the form 
\begin{eqnarray}
{\cal L}_{\rm G} = \frac{1}{4 g_{\rm G}^{2} } \tr ({F}^{\rm G}_{\mu \nu})^2
+ 
\sum_{f=1}^{N_f} 
\bar{\psi}^{\rm G}_{f }\left( \gamma^{\mu} {D}^G_{\mu} + m_f + \mu_f \gamma^4 \right)\psi^{\rm G}_{f }, 
\label{QCDlike_action}
\end{eqnarray}
where $G$ is the gauge group e.g. $SU(N_c)$, 
$\mu_f$ is a generic quark chemical potential, and fermions $\psi^{\rm G}$ are not necessarily in the fundamental representation. 
The main examples are QCD with an isospin chemical potential $\mu_I$ (i.e. $N_f=2$, $\mu_1=-\mu_2=\mu_I/2$) 
and degenerate mass $m_1=m_2$, 
two-color QCD of even number of flavors and degenerate mass, $SU(N_c)$ Yang-Mills  with adjoint fermions, 
and $SO(2N_c)$ and $Sp(2N_c)$ Yang-Mills theories.  
However, these theories look quite different from $N_{c}=3$ QCD;  
for example the flavor symmetry is explicitly broken in the first case. Therefore it is important to understand  
{\it what we can learn from these theories}, or in other words, {\it in what sense they are similar to real QCD with the baryon chemical potential}.  

\begin{figure}[htb]
\begin{center}
\epsfig{file=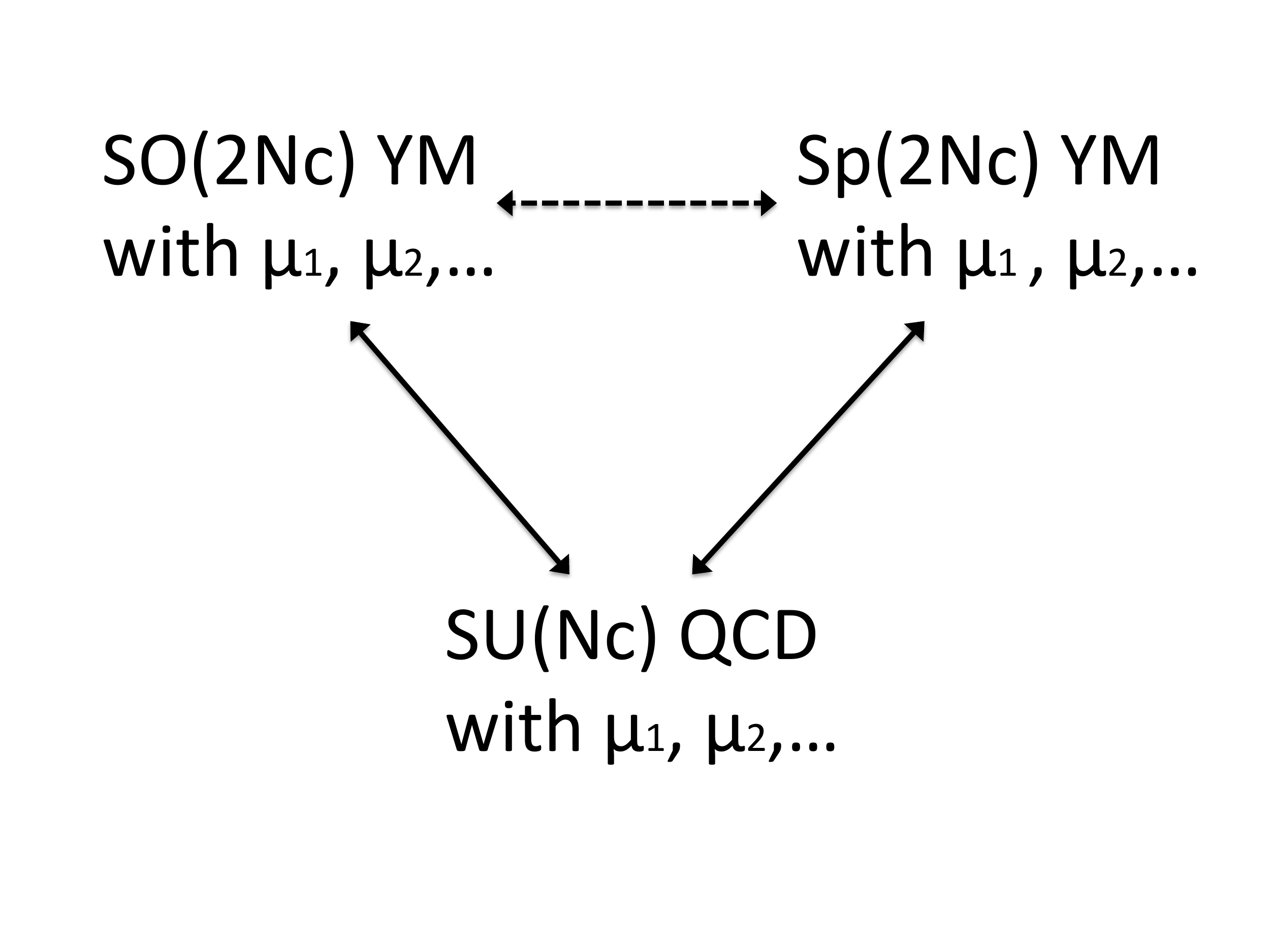,height=1.5in}
\caption{More general version of the equivalences. 
Values of the quark chemical potentials can be different. }
\label{fig:web_general}
\end{center}
\end{figure}

In \cite{Cherman:2010jj,Hanada:2011ju,Hanada:2011ev}, 
an answer to this question has been given. 
The statements are 
\begin{itemize}
\item

$SO(2N_c)$ YM with $\mu_B$, $Sp(2N_c)$ YM with $\mu_B$ and $SU(N_c)$ QCD with $\mu_I$ are large-$N_c$ equivalent 
both in the 't Hooft limit ($N_f$ fixed) and the Veneziano limit ($N_f/N_c$ fixed), everywhere in the $T$-$\mu$ plane. 
(Fermions are in the fundamental (vector) representations.)

\item
$SO(2N_c)$ YM with $\mu_B$, $Sp(2N_c)$ YM with $\mu_B$, $SU(N_c)$ QCD with $\mu_I$ 
and $SU(N_c)$ QCD with $\mu_B$ are large-$N_c$ equivalent 
in the 't Hooft limit, outside the BEC/BCS crossover region of the former three theories. 
(Fermions are in the fundamental representations.)

\item
More generally, $SO(2N_c)$, $Sp(2N_c)$ and $SU(N_c)$ theories with fermion mass $m_1,\cdots,m_{N_f}$ and 
chemical potential $\mu_1,\cdots,\mu_{N_f}$ are equivalent. The signs of the chemical potential can be flipped 
without spoiling the equivalence. (Fig.~\ref{fig:web_general}) 

\item
$SO(2N_c)$ YM with the $N_f$  complex adjoint fermions and $\mu_B$, 
$SU(N_c)$ YM with the $N_f$  complex adjoint fermions and $\mu_B$, 
and $SU(N_c)$ YM with the $2N_f$  complex anti-symmetric fermions and $\mu_I$ 
are large-$N_c$ equivalent  everywhere in the $T$-$\mu$ plane. 

\end{itemize}

\begin{figure}[htb]
\begin{center}
\epsfig{file=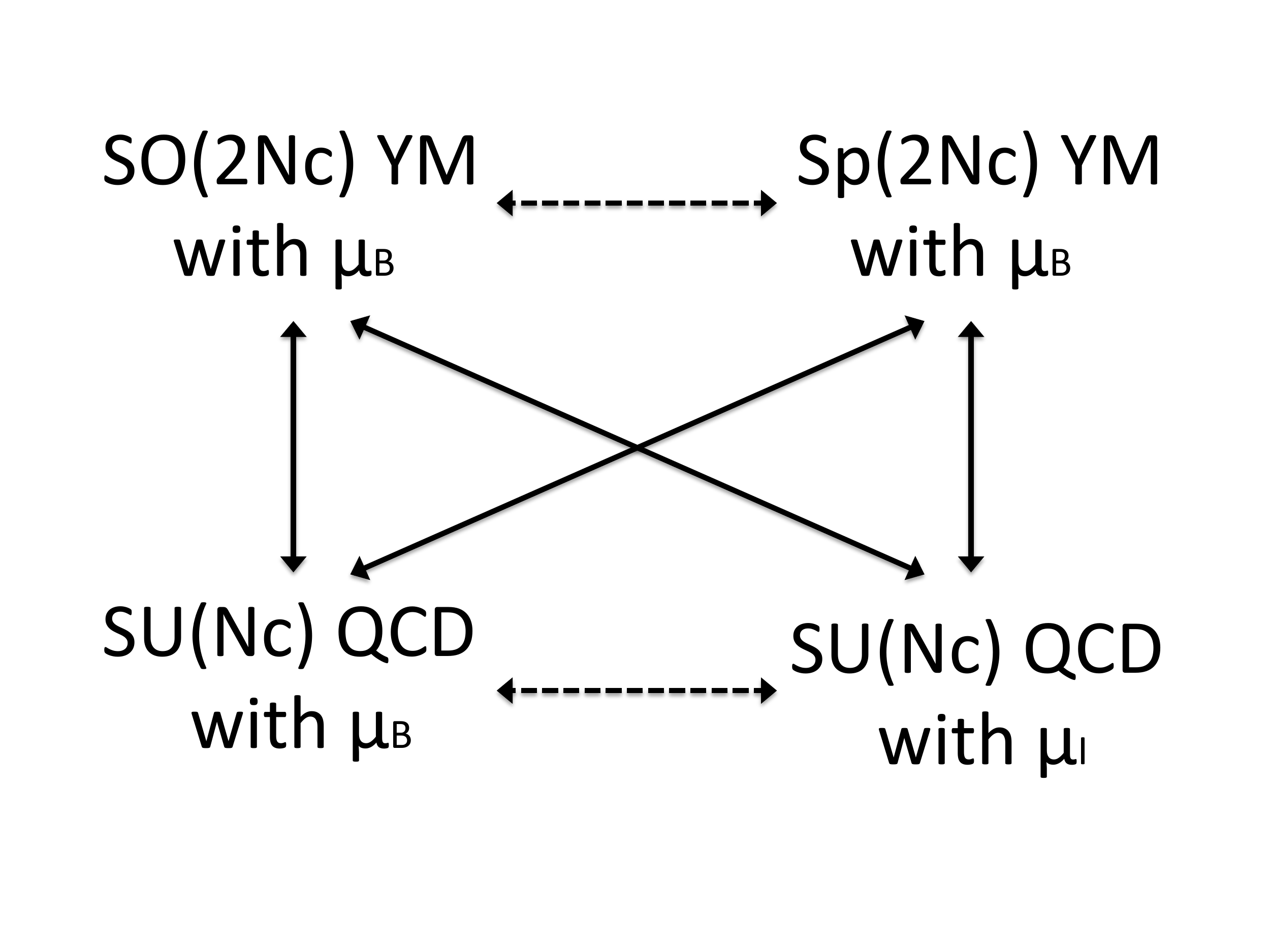,height=1.5in}
\caption{A web of equivalences. Arrows with solid lines represent equivalences through orbifold projections. 
Arrows with dashed lines are the `parent-parent' and `daughter-daughter' equivalences which arise as combinations 
of two orbifold equivalences.   }
\label{fig:web}
\end{center}
\end{figure}

\begin{figure}[htb]
\begin{center}
\epsfig{file=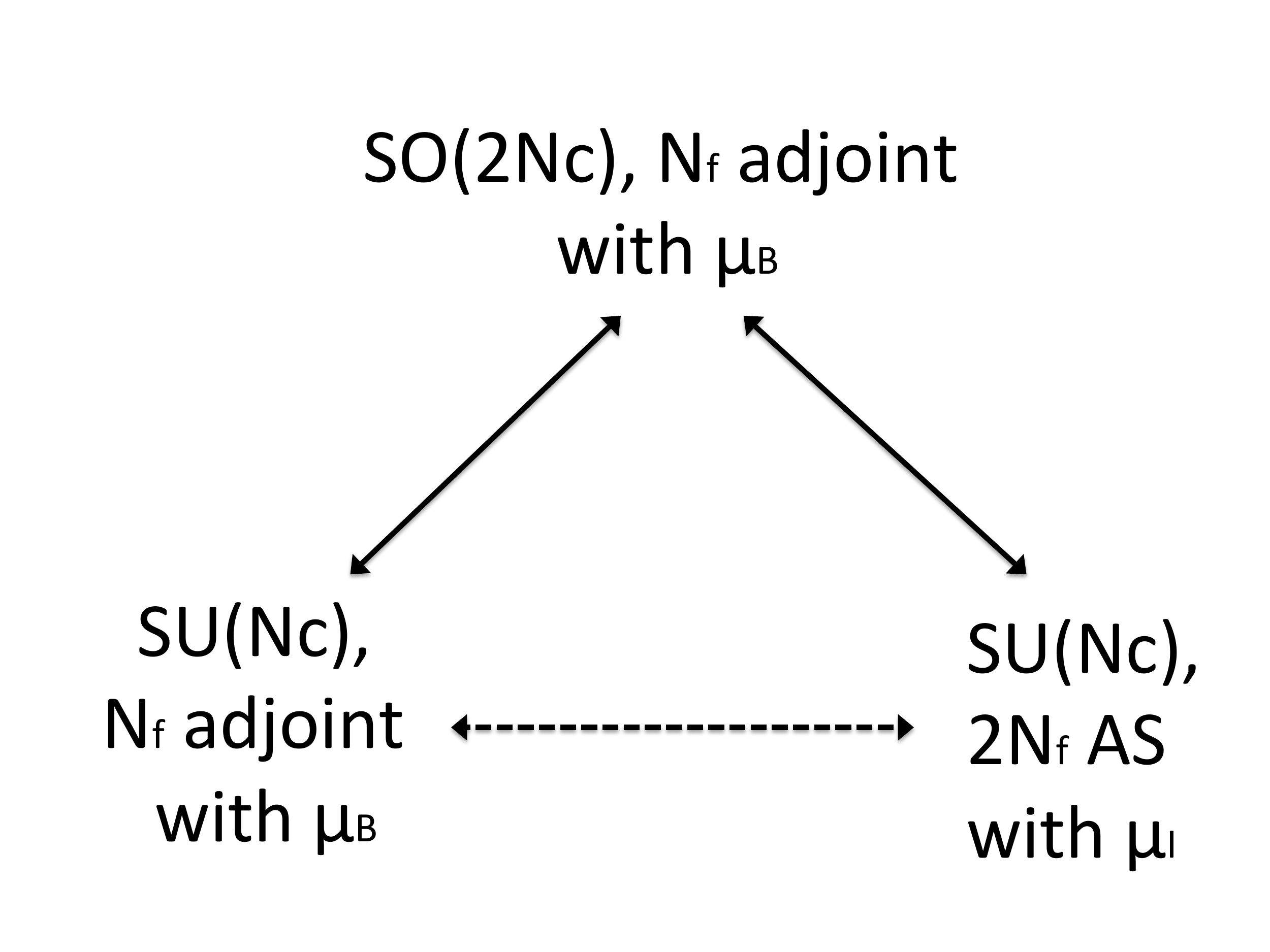,height=1.5in}
\caption{Equivalences in the Corrigan-Ramond limit. $SU(N_c)$ YM with anti-symmetric fermions 
can be regarded as a special kind of large-$N_c$ limit of three-color QCD (the Corrigan-Ramond limit), because 
anti-symmetric and fundamental representations are equivalent at $N_c=3$. 
Unfortunately, $SU(N_c)$ YM with anti-symmetric fermions and $\mu_B$ 
cannot be incorporated in these equivalences. }
\label{fig:web2}
\end{center}
\end{figure}

These statements have been derived by using a string-inspired large-$N_c$ technique,  
which is called the {\it orbifold equivalence} \cite{Kachru:1998ys,Bershadsky:1998mb,Bershadsky:1998cb,Kovtun:2003hr}.
As shown in \cite{Cherman:2010jj,Hanada:2011ju,Hanada:2011ev}, there are orbifold projections relating 
$SO(2N_c)$ and $Sp(2N_c)$ theories with $\mu_B$, 
QCD with $\mu_B$ and QCD with $\mu_I$ (Fig.~\ref{fig:web}). 
At large-$N_c$, the orbifold equivalence 
guarantees these theories are equivalent in the sense a class of correlation functions (e.g. the expectation value of the chiral condensate and 
$\pi^0$ correlation functions) 
and the phase diagrams determined by such quantities coincide, 
as long as the projection symmetry is not broken spontaneously \cite{Kovtun:2003hr}. 
A similar argument shows QCD with adjoint fermions and $\mu_B$ 
is equivalent to QCD with fermions of two-index antisymmetric representation, which is the so-called Corrigan-Ramond large-$N_c$ limit, 
with $\mu_I$ (Fig.~\ref{fig:web2}).  
In order for these equivalences to hold, orbifolding symmetries must not be broken spontaneously. 
This requirement is always satisfied for the equivalences between 
$SO(2N_c)$ YM with $\mu_B$, $Sp(2N_c)$ YM with $\mu_B$ and QCD with $\mu_I$. 
For the equivalences between these three theories and QCD with the baryon chemical potential,  
`outside the BEC/BCS crossover region' is required for the symmetry realization. 
This region is relevant for the search for the QCD critical point, 
which attract intense interest over the decade.  
Our answer to the problem is strikingly simple -- one can study it by using the sign-free theories.  
In the case of the two-flavor theory, QCD with $\mu_I$ is nothing but the phase-quenched version of QCD with $\mu_B$. 
Therefore, the sign problem is merely an illusion, up to the $1/N_c$ correction. 
Furthermore, for gluonic observables, the leading $1/N_c$ corrections to the large-$N_c$ limit which carry the information of the chemical potential  
are the same in these theories.

%%%%%%%%%%%%%%%%%%%%%%%%%%%%%%%%%%%%%%%%%%%%%%%%%%%%%%%%%%%%%%%%%%%%%%%%%
%%
%%   use this format to include an .eps figure into your paper
%%
%%%%%%%%%%%%%%%%%%%%%%%%%%%%%%%%%%%%%%%%%%%%%%%%%%%%%%%%%%%%%%%%%%%%%%%%%%%

%%%%%%%%%%%%%%%%%%%%%%%%%%%%%%%%%%%%%%%%%%%%%%%%%%%%%%%%%%%%%%%%%%%%%%%%%
%%
%%   use this format to include an .eps figure into your paper
%%
%%%%%%%%%%%%%%%%%%%%%%%%%%%%%%%%%%%%%%%%%%%%%%%%%%%%%%%%%%%%%%%%%%%%%%%%%%%

%%%%%%%%%%%%%%%%%%%%%%%%%%%%%%%%%%%%%%%%%%%%%%%%%%%%%%%%%%%%%%%%%%%%%%%%%
%%
%%   use this format to include an .eps figure into your paper
%%
%%%%%%%%%%%%%%%%%%%%%%%%%%%%%%%%%%%%%%%%%%%%%%%%%%%%%%%%%%%%%%%%%%%%%%%%%%%

%%%%%%%%%%%%%%%%%%%%%%%%%%%%%%%%%%%%%%%%%%%%%%%%%%%%%%%%%%%%%%%%%%%%
%%%%%%%%%%%%%%%%%%%%%%%%%%%%%%%%%%%%%%%%%%%%%%%%%%%%%%%%%%%%%%%%%%%%
%%%%%%%%%%%%%%%%%%%%%%%%%%%%%%%%%%%%%%%%%%%%%%%%%%%%%%%%%%%%%%%%%%%%
\section{Orbifold equivalence}\label{sec:equivalence}
%%%%%%%%%%%%%%%%%%%%%%%%%%%%%%%%%%%%%%%%%%%%%%%%%%%%%%%%%%%%%%%%%%%%
%%%%%%%%%%%%%%%%%%%%%%%%%%%%%%%%%%%%%%%%%%%%%%%%%%%%%%%%%%%%%%%%%%%%
%%%%%%%%%%%%%%%%%%%%%%%%%%%%%%%%%%%%%%%%%%%%%%%%%%%%%%%%%%%%%%%%%%%% 
%
%%%%%%%%%%%%%%%%%%%%%%%%%%%%%%%%%%%%%%%%%%%%%%%%%%%%%%%%%%%%%%%%%%%%
%%%%%%%%%%%%%%%%%%%%%%%%%%%%%%%%%%%%%%%%%%%%%%%%%%%%%%%%%%%%%%%%%%%%
%%%%%%%%%%%%%%%%%%%%%%%%%%%%%%%%%%%%%%%%%%%%%%%%%%%%%%%%%%%%%%%%%%%%
\subsection{Pure Yang-Mills theory}\label{proof:pureYM}
%%%%%%%%%%%%%%%%%%%%%%%%%%%%%%%%%%%%%%%%%%%%%%%%%%%%%%%%%%%%%%%%%%%%
%%%%%%%%%%%%%%%%%%%%%%%%%%%%%%%%%%%%%%%%%%%%%%%%%%%%%%%%%%%%%%%%%%%%
%%%%%%%%%%%%%%%%%%%%%%%%%%%%%%%%%%%%%%%%%%%%%%%%%%%%%%%%%%%%%%%%%%%% 
The notion of the orbifold equivalence came from the string theory \cite{Kachru:1998ys,Bershadsky:1998mb}. 
Soon it was proven by using only field theory techniques 
\cite{Bershadsky:1998cb,Kovtun:2003hr}, without referring to the string theory.  
As a simple example, let us consider the equivalence between $SO(2N_c)$ and $SU(N_c)$ pure Yang-Mills theories. 
(The projection from $Sp(2N_c)$ to $SU(N_c)$ is obtained in a similar manner.)
To perform an orbifold projection, one identifies a discrete subgroup of the symmetry group of the `parent' theory, 
which is the $SO(2N_{c})$ theory in this case, and requires the fields to be invariant under the discrete symmetry.  
This gives a `daughter' theory, which is the $SU(N_{c})$ YM.  

The details are as follows. 
Let us take $J_c\in SO(2N_{c})$ to be $J_c = i\sigma_{2} \otimes 1_{N_{c}}$, which generates a $\mathbb{Z}_{4}$ subgroup of $SO(2N_{c})$. 
Here $1_{N}$ is an $N \times N$ identity matrix.    
We require the gauge field $A_\mu$ to be invariant under 
\begin{eqnarray}
A_{\mu} \rightarrow J_c A_{\mu} J_c^{-1},  
\end{eqnarray}
which generates a $\mathbb{Z}_{2}$ subgroup of $SO(2N_{c})$. 
A generic $SO(2N_c)$ gauge field $A_{\mu}$ can be written in $N_{c}\times N_{c}$ blocks as
\begin{align}
A_\mu
=
i\left(
\begin{array}{cc}
A_\mu^A+B_\mu^A & C_\mu^A-D_\mu^S\\
C_\mu^A+D_\mu^S & A_\mu^A-B_\mu^A
\end{array}
\right),
\end{align}
where fields with an `$A$' (`$S$') superscript are anti-symmetric (symmetric) matrices.  
Under the $\mathbb{Z}_{2}$ symmetry, $A_\mu^A$ and $D_\mu^S$ are even while $B_\mu^A$ and $C_\mu^A$ are odd,  
and hence the orbifold projection sets $B_{\mu}^{A} = C_{\mu}^{A} = 0$;   
the `daughter' field is  
\begin{align}
A_{\mu}^{\rm proj}
=
i\left(
\begin{array}{cc}
A_{\mu}^A  & -D_\mu^S\\
D_\mu^S & A_\mu^A
\end{array}
\right).
\end{align} 
By using a unitary matrix
\begin{eqnarray}
P = \frac{1}{\sqrt{2}}\left(
\begin{array}{cc}
1_{N_{c}} & i 1_{N_{c}} \\
1_{N_{c}} & -i 1_{N_{c}}
\end{array} 
\right),
\end{eqnarray}
it can be rewritten as 
\begin{eqnarray}
P A_{\mu}^{\rm proj} P^{-1} =
  \left(
\begin{array}{cc}
-\mathcal{A}_{\mu}^{T} & 0\\
0 & \mathcal{A}_{\mu}
\end{array} 
\right),
\end{eqnarray}
where $\mathcal{A}_{\mu} \equiv D_{\mu}^{S} + i A^{A}_{\mu}$ is a $U(N_{c})$ gauge field.  
However, the difference between $U(N_{c})$ and $SU(N_{c})$ is a $1/N_{c}^{2}$ correction and can be neglected at large-$N_c$.~\footnote{
When one studies $U(N_c)$ theory, it is difficult to control the $U(1)$ part and the effect of the chemical potential can be washed away 
by a lattice artifact. In order to avoid it one should simulate $SU(N_c)$ theory on the lattice. 
} The gauge part of the action after the orbifold projection is thus simply
\begin{eqnarray}
\mathcal{L^{\mathrm{gauge}, \mathrm{proj}}} = \frac{2}{4g_{SO}^{2}} Tr \mathcal{F}_{\mu \nu}\mathcal{F}^{\mu \nu}, 
\end{eqnarray}
where $\mathcal{F}_{\mu\nu}$ is the $SU(N_{c})$ field strength. 
Let us identify it with the Lagrangian of the daughter theory times two, 
\begin{eqnarray}
\label{eq:recipe}
{\cal L}_{\rm SO} \rightarrow 2 {\cal L}_{\rm SU},
\end{eqnarray}
or equivalently  
let us take  $g_{SU}^{2} = g_{SO}^{2}$, where $g_{SU}$ is the gauge coupling constant of the $SU(N_{c})$ theory. 
This factor two is necessary in order for the ground state energies, which are proportional to the degrees of freedom, to match. 
Given this projection, expectation values of the gauge-invariant operators in parent theory, ${\cal O}^{(p)}[A_\mu]$, agree with the expectation values of 
counterparts in the daughter theory, which are obtained by replacing $A_\mu$ with $\cal A_\mu^{\rm proj}$, %that is,  
${\cal O}^{(d)}[{\cal A}_\mu]\equiv {\cal O}^{(p)}[A_\mu^{\rm proj}]$ in the large-$N_c$ limit \cite{Bershadsky:1998cb,Kovtun:2003hr}, 
as long as the projection symmetry is not broken spontaneously.

%%%%%%%%%%%%%%%%%%%%%%%%%%%%%%%%%%%%%%%%%%%%%%%%%%%%%%%%%%%%%%%%%%%%
%%%%%%%%%%%%%%%%%%%%%%%%%%%%%%%%%%%%%%%%%%%%%%%%%%%%%%%%%%%%%%%%%%%%
%%%%%%%%%%%%%%%%%%%%%%%%%%%%%%%%%%%%%%%%%%%%%%%%%%%%%%%%%%%%%%%%%%%%
\subsection{Introducing fundamental fermions}
\hspace{0.51cm}
%%%%%%%%%%%%%%%%%%%%%%%%%%%%%%%%%%%%%%%%%%%%%%%%%%%%%%%%%%%%%%%%%%%%
%%%%%%%%%%%%%%%%%%%%%%%%%%%%%%%%%%%%%%%%%%%%%%%%%%%%%%%%%%%%%%%%%%%%
%%%%%%%%%%%%%%%%%%%%%%%%%%%%%%%%%%%%%%%%%%%%%%%%%%%%%%%%%%%%%%%%%%%%
\subsubsection{Orbifold projection of fundamental fermions}
\hspace{0.51cm}
%%%%%%%%%%%%%%%%%%%%%%%%%%%%%%%%%%%%%%%%%%%%%%%%%%%%%%%%%%%%%%%%%%%%
%%%%%%%%%%%%%%%%%%%%%%%%%%%%%%%%%%%%%%%%%%%%%%%%%%%%%%%%%%%%%%%%%%%%
%%%%%%%%%%%%%%%%%%%%%%%%%%%%%%%%%%%%%%%%%%%%%%%%%%%%%%%%%%%%%%%%%%%% 
In this section, we introduce the orbifold projection for fundamental fermions \cite{Cherman:2010jj,Hanada:2011ju}. 
By using $\omega=e^{i\pi/2}\in U(1)_B$, we define the projection by 
\begin{eqnarray}
\psi_f = \omega J_c \psi_f. 
\label{baryon_projection}
\end{eqnarray}
By using 
 \begin{eqnarray}
\left(
\begin{array}{c}
\psi^{+}_{f}\\
\psi^{-}_{f}
\end{array}
\right)
\equiv
P\psi_{f}, 
\end{eqnarray}
the action of the $\mathbb{Z}_{2}$ symmetry is just  $(\psi^{+}_{f}, \psi^{-}_{f})\rightarrow (- \psi^{+}_{f}, \psi^{-}_{f})$. The projection consists of setting $\psi^{+}_{f} = 0$.

The action of the daughter theory is 
\begin{eqnarray}
\mathcal{L} = \frac{1}{4 g_{SU}^{2} } Tr \mathcal{F}_{\mu \nu}^2
+ 
\sum_{f=1}^{N_f}
\bar{\psi}_{f}^{\rm SU}\left( \gamma^{\mu} {\cal D}_{\mu} + m_q + \mu\gamma^4\right)\psi_{f}^{\rm SU}, 
\end{eqnarray}
where $\mathcal{F}_{\mu\nu}$ is the field strength of the $SU(N_{c})$ gauge field $\mathcal{A}_{\mu} = D^{S}_{\mu} + i A^{A}_{\mu}$, 
$\psi_{f}^{\rm SU} = \psi^{-}_{f}$, and ${\cal D}_{\mu} = \partial_{\mu} + i \mathcal{A}_{\mu}$.  
This is an $SU(N_{c})$ gauge theory with $N_{f}$ flavors of fundamental Dirac fermions 
and the baryon chemical potential $\mu_B=\mu N_c$.   So the orbifold projection relates $SO(2N_{c})$ gauge theory to large $N_{c}$ QCD.  

On the other hand, in order to obtain fermions at finite $\mu_I$ for even $N_f$, 
we use $J_c \in SO(2N_c)$ [or $J_c \in Sp(2N_c)$] and 
$J_i \in SU(2)_{\rm isospin} [\subset SU(N_f)]$ defined by
\begin{eqnarray}
\label{eq:J_i}
J_i = - i\sigma_2 \otimes 1_{N_f/2}.
\end{eqnarray}
We choose the projection condition to be
\begin{eqnarray}
\label{eq:projection_isospin}
(J_c)_{aa'} \psi_{a'f'} (J_i^{-1})_{f'f}=\psi_{af}. 
\end{eqnarray}
The flavor $N_f$-component fundamental fermion is decomposed into 
two $(N_f/2)$-component fields, 
\begin{eqnarray}
\psi
=
(\psi_i \ \psi_j), 
\end{eqnarray}
with $i$ and $j$ being the isospin indices.
If we define $\varphi_{\pm}=(\psi_{\pm}^i \mp i \psi_{\pm}^j)/\sqrt{2}$
and $\xi_{\pm}=(\psi_{\pm}^i \pm \ i \psi_{\pm}^j)/\sqrt{2}$,
$\varphi_{\pm}$ survive but $\xi_{\pm}$ disappear after the projection
(\ref{eq:projection_isospin}).
Because $\varphi_{\pm}$ couple to $(A_{\mu}^{\rm SU})^C$ and $A_{\mu}^{\rm SU}$ respectively, 
the action of the daughter theory is expressed as
\begin{eqnarray}
{\cal L}_{\rm SU} = \frac{1}{4 g_{\rm SU}^{2} } \tr ({F}^{\rm SU}_{\mu \nu})^2
+ 
\sum_{f=1}^{N_f/2} \sum_{\pm}
\bar{\psi}^{\rm SU}_{f \pm}\left( \gamma^{\mu} {D}_{\mu} + m \pm \mu \gamma^4 \right)\psi^{\rm SU}_{f \pm}, 
\end{eqnarray}
where $\psi^{\rm SU}_{+}\equiv\sqrt{2}\varphi_-$ and $\psi^{\rm SU}_{-}\equiv\sqrt{2}\varphi_+^C$.
This theory has the isospin chemical potential $\mu_I=2\mu$.

%%%%%%%%%%%%%%%%%%%%%%%%%%%%%%%%%%%%%%%%%%%%%%%%%%%%%%%%%%%%%%%%%%%%
%%%%%%%%%%%%%%%%%%%%%%%%%%%%%%%%%%%%%%%%%%%%%%%%%%%%%%%%%%%%%%%%%%%%
%%%%%%%%%%%%%%%%%%%%%%%%%%%%%%%%%%%%%%%%%%%%%%%%%%%%%%%%%%%%%%%%%%%%
\subsubsection{The 't Hooft limit vs the Veneziano limit}\label{'tHooftVsVeneziano}
\hspace{0.51cm}
%%%%%%%%%%%%%%%%%%%%%%%%%%%%%%%%%%%%%%%%%%%%%%%%%%%%%%%%%%%%%%%%%%%%
%%%%%%%%%%%%%%%%%%%%%%%%%%%%%%%%%%%%%%%%%%%%%%%%%%%%%%%%%%%%%%%%%%%%
%%%%%%%%%%%%%%%%%%%%%%%%%%%%%%%%%%%%%%%%%%%%%%%%%%%%%%%%%%%%%%%%%%%%
The proof of the orbifold equivalence of the pure Yang-Mills theories in \cite{Bershadsky:1998cb} 
can be applied even with the fundamental fermion, when the chemical potential is zero. 
Note that two projections \eqref{baryon_projection} and 
\eqref{eq:projection_isospin} are equivalent when the chemical potential is absent. 
Both are a $\mathbb{Z}_4$ subgroup of the flavor symmetry which mixes two Majorana flavors. 
Once the chemical potential is turned on, they are not equivalent. 
The flavor symmetry $J_i$ used in \eqref{eq:projection_isospin} satisfies the assumption  
used in \cite{Bershadsky:1998cb}, and the proof can be repeated straightforwardly. 
On the other hand, $\mathbb{Z}_4\in U(1)_B$ used in  \eqref{baryon_projection} does not satisfy that assumption; 
however it is still possible to show that all planar diagrams with at most one fermion loop coincide. 
Because the fermion loops are suppressed by the factor $N_f/N_c$, the equivalence through \eqref{baryon_projection} holds 
in the 't Hooft large-$N_c$ limit ($N_f$ fixed) while the one through \eqref{eq:projection_isospin} holds also in the Veneziano limit ($N_f/N_c$ fixed). 

The above argument has an implication for the $1/N_c$ correction. Let us consider QCD with $\mu_B$ and that with $\mu_I$. 
In the 't Hooft large-$N_c$ limit, gluonic operators trivially agree because the fermions are not dynamical. 
Let us consider finite-$N_c$, say $N_c=3$ and $N_f=2$. Then the largest correction to the 't Hooft limit comes from 
one-fermion-loop planar diagrams, which, as we have seen,  do not distinguish $\mu_B$ and $\mu_I$. 
Therefore gluonic operators should behave similarly even quantitatively; the difference is at most $(N_f/N_c)^2$ (two-fermion-loop planar diagrams) 
or $(1/N_c^2)\cdot(N_f/N_c)$ (one-fermion-loop nonplanar diagrams). In particular, the deconfinement temperatures, which are determined from 
the Polyakov loop, should be close. 
%%%%%%%%%%%%%%%%%%%%%%%%%%%%%%%%%%%%%%%%%%%%%%%%%%%%%%%%%%%%%%%%%%%%
%%%%%%%%%%%%%%%%%%%%%%%%%%%%%%%%%%%%%%%%%%%%%%%%%%%%%%%%%%%%%%%%%%%%
%%%%%%%%%%%%%%%%%%%%%%%%%%%%%%%%%%%%%%%%%%%%%%%%%%%%%%%%%%%%%%%%%%%%
\subsubsection{Symmetry realization and validity of the equivalence}\label{sec:phase}
\hspace{0.51cm}
%%%%%%%%%%%%%%%%%%%%%%%%%%%%%%%%%%%%%%%%%%%%%%%%%%%%%%%%%%%%%%%%%%%%
%%%%%%%%%%%%%%%%%%%%%%%%%%%%%%%%%%%%%%%%%%%%%%%%%%%%%%%%%%%%%%%%%%%%
%%%%%%%%%%%%%%%%%%%%%%%%%%%%%%%%%%%%%%%%%%%%%%%%%%%%%%%%%%%%%%%%%%%% 
As we have seen so far, $SU(N_c)$ QCD with $\mu_B/\mu_I$, $SO(2N_c)$ YM and $Sp(2N_c)$ YM should be  
equivalent in the large-$N_c$ limit as long as the projection symmetries are not broken spontaneously. 
In this section we discuss the phase structures of these theories and clarify when the symmetries are broken.  
It turns out that $SU(N_c)$ QCD with $\mu_I$, $SO(2N_c)$ YM and $Sp(2N_c)$ YM  with $\mu_B$ should be equivalent 
everywhere in $T$-$\mu$ parameter space. 
The equivalence to  $SU(N_c)$ QCD with $\mu_B$ is not valid outside the BEC/BCS crossover region of other three theories. 
(In \cite{Cherman:2010jj,Cherman:2011mh} a possible cure to this is discussed.) 

Let us start with $SO(2N_c)$ YM with $\mu_B$.  
A crucial difference from QCD is that there is no distinction between `matter' and `antimatter' because the gauge group is real.  
In other words, `fundamental' and `anti-fundamental' representations are equivalent. 
For this reason, mesons in this theory are not necessarily neutral under $U(1)_B$; one can construct `baryonic mesons' and 
`anti-baryonic mesons' out of two `quarks' and `antiquarks', respectively.  
Because they couple to $\mu_B$, 
as we increase the value of $\mu_B$ the lightest `baryonic meson' condenses at some point.  
Then the $U(1)_B$ symmetry is broken to ${\mathbb Z}_2$ and the equivalence to QCD with $\mu_B$ fails. 
(Note that we have used ${\mathbb Z}_4$ subgroup of $U(1)_B$ for the projection.)

\begin{figure}[t]
\begin{center}
\includegraphics[width=10cm]{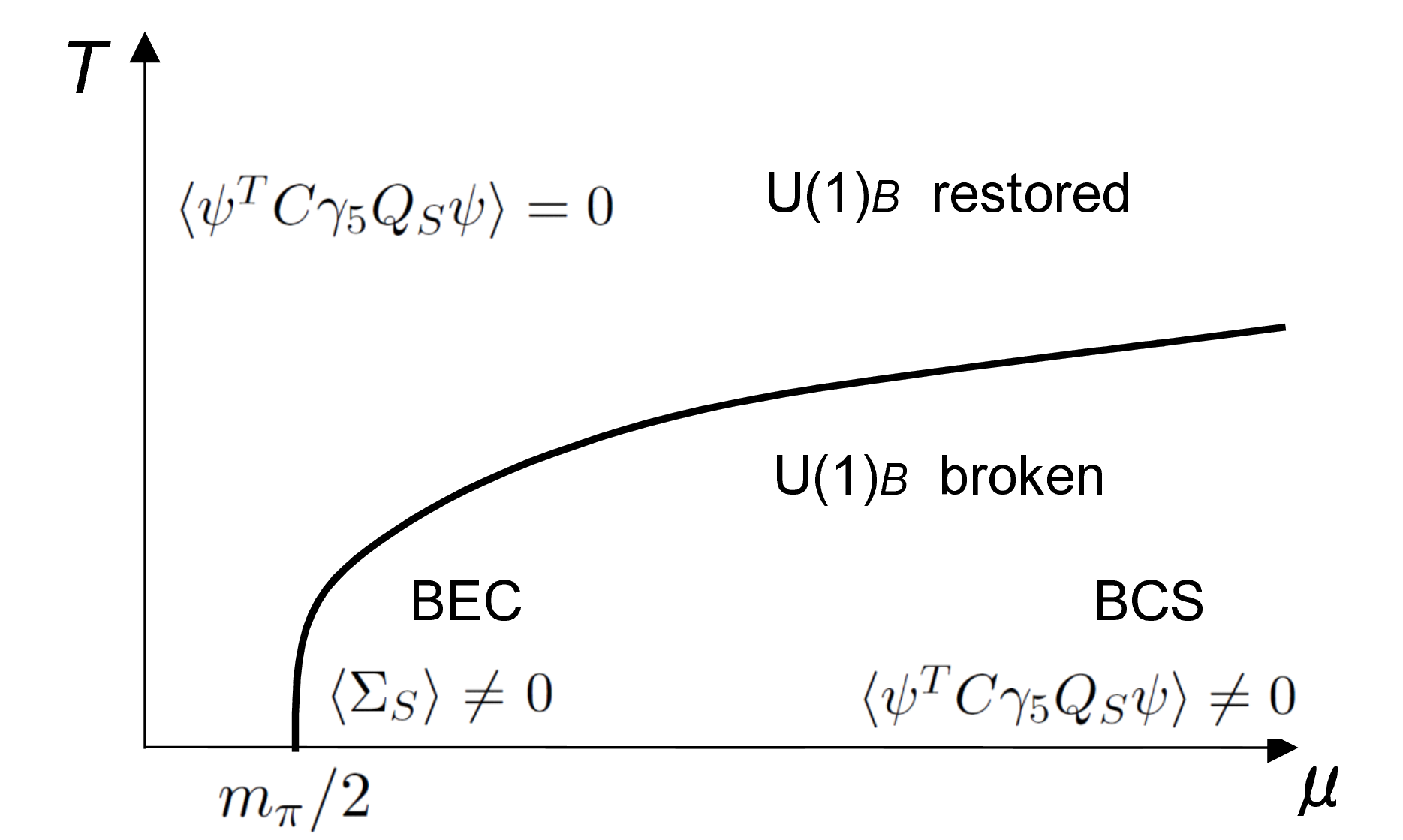}
\end{center}
\vspace{-0.5cm}
\caption{Phase diagram of $SO(2N_c)$ gauge theory at finite $\mu_B$. 
(Figure taken from \cite{Hanada:2011ju}.)}
\label{fig:SO}
\end{figure}

\begin{figure}[t]
\begin{center}
\includegraphics[width=10cm]{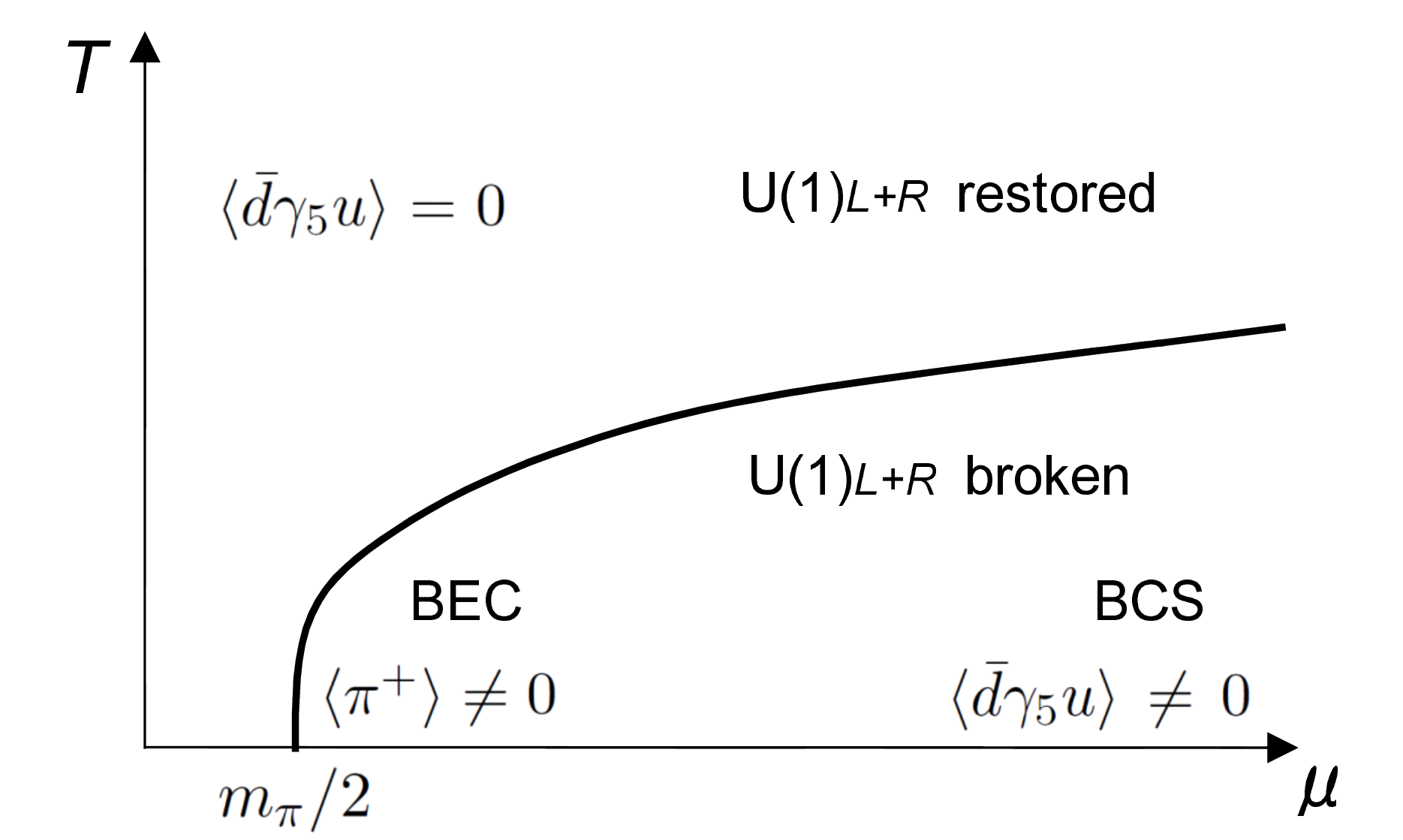}
\end{center}
\vspace{-0.5cm}
\caption{Phase diagram of QCD at finite $\mu_I=2\mu$. 
(Figure taken from \cite{Hanada:2011ju}.)}
\label{fig:muI}
\end{figure}

In order to identify the lightest baryonic meson, let us consider the chiral symmetry breaking in this theory. 
When $m = \mu_B = 0$, the Lagrangian \eqref{QCDlike_action} has the
$SU(N_{f})_{L}\times SU(N_{f})_{R} \times U(1)_{B} \times U(1)_{A}$ 
symmetry at the classical level at first sight.  
However, chiral symmetry of the theory is known to be enhanced to $SU(2N_{f})$. 
Here $U(1)_A$ is explicitly broken by the axial anomaly.
One can actually rewrite the fermionic part of the Lagrangian (\ref{QCDlike_action})
manifestly invariant under $SU(2N_f)$
using the new variable $\Psi=(\psi_L, \sigma_2 \psi_R^*)^T$:
\begin{eqnarray}
{\cal L}_{\rm f}=i \Psi^{\dag} \sigma_{\mu} D_{\mu} \Psi,
\end{eqnarray}
where $\sigma_{\mu}=(\sigma_k,-i)$ with the Pauli matrices $\sigma_k$ ($k=1,2,3$).
The chiral symmetry $SU(2N_f)$ is spontaneously broken down to 
$SO(2N_{f})$ by the formation of the chiral condensate 
$\langle \bar{\psi}{\psi} \rangle$, leading to the $2N_f^2 + N_f -1$ Nambu-Goldstone bosons 
living on the coset space $SU(2N_{f})/SO(2N_{f})$ : 
neutral pions $\Pi_a=\bar{\psi} \gamma_{5} P_a \psi$, 
`baryonic pions' (or simply `diquark')
$\Sigma_S = \psi^{T} C \gamma_5 Q_S \psi$ and 
`anti-baryonic pions' $\Sigma_S^{\dag}= \psi^{\dag} C \gamma_5 Q_S \psi^*$.
It is easy to see the fate of these bosons under the orbifold projection. 
The projection to QCD with $\mu_B$ maps  neutral pions to pions in QCD, and baryonic and anti-baryonic pions are projected away. 
On the other hand, the projection to QCD with $\mu_I$ sends neutral/baryonic/anti-baryonic pions to 
$\pi^0$, $\pi^+$ and $\pi^-$, respectively. 
Therefore the (baryonic) pions in $SO(2N_c)$ YM and those in QCD have the same mass $m_{\pi}$. 
In the same way as the $\pi^+$ condensation in QCD with $\mu_I$ at $\mu=m_\pi/2$, 
baryonic pions condense at $\mu=m_\pi/2$ (Fig.~\ref{fig:SO} and Fig.~\ref{fig:muI}).

%%%%%%%%%%%%%%%%%%%%%% 
%%%%%%%%%%%%%%%%%%%%%%%%%

%%%%%%%%%%%%%%%%%%%%%% 
%%%%%%%%%%%%%%%%%%%%%%%%%

At sufficiently large $\mu$, 
the one-gluon exchange interaction in the $\psi \psi$-channel
is attractive in the color symmetric channel,
leading to the condensation of the diquark pairing $\langle \psi^{T} C \gamma_5 Q_S \psi \rangle$.
This diquark condensate does not break $SO(2N_c)$ symmetry.
This BCS pairing has the same quantum numbers 
and breaks the same $U(1)_B$ symmetry as the BEC $\langle \Sigma_S \rangle $ 
at small $\mu_B$, and there should be no phase transition for 
$\mu>m_{\pi}/2$ along $\mu$ axis.
The phase diagram of this theory is similar to that of 
QCD at finite $\mu_I$, as shown in Fig.~\ref{fig:SO} and Fig.~\ref{fig:muI}. 
This is because the condensates in two theories are related each other 
through the orbifold projection, and furthermore, the condensation does not break the flavor symmetry 
used for the projection.   

\begin{figure}[t]
\begin{center}
\includegraphics[width=10cm]{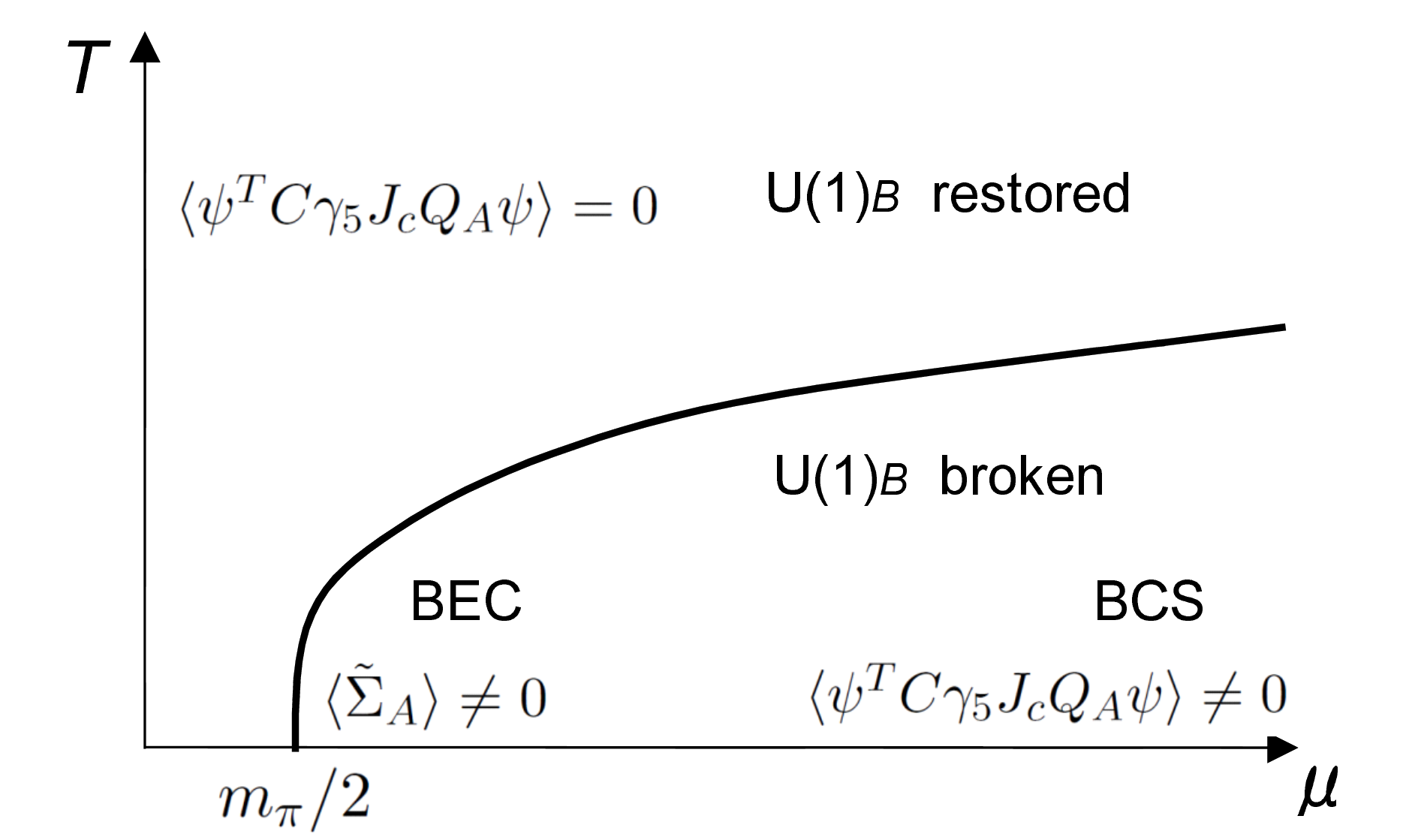}
\end{center}
\vspace{-0.5cm}
\caption{Phase diagram of $Sp(2N_c)$ gauge theory at finite $\mu_B$. 
$\tilde \Sigma_A = \psi^{T} C \gamma_5 J_c Q_A \psi$, where 
$Q_A$ ($A=1,2,\cdots,N_f(N_f - 1)/2$) are antisymmetric $N_f \times N_f$ matrices  in the flavor space. 
(Figure taken from \cite{Hanada:2011ju}.)
}
\label{fig:Sp}
\end{figure}

In the same manner, $Sp(2N_c)$ YM and QCD with $\mu_I$ are equivalent everywhere in $T$-$\mu$ plane; 
see Fig.~\ref{fig:Sp}. (For further details, see \cite{Hanada:2011ju}.) 

%%%%%%%%%%%%%%%%%%%%%% 
%%%%%%%%%%%%%%%%%%%%%%%%%

QCD with $\mu_B$ behaves rather differently, because $\mu_B$ does not couple to mesons. 
This does not lead to a contradiction, however. Because baryons are much heavier than pions, 
phenomena characteristic to QCD with $\mu_B$ (e.g. formation of hadronic matter) takes place 
after the equivalence is gone due to the $U(1)_B$ breakdowns in $SO(2N_c)$ and $Sp(2N_c)$ Yang-Mills. 

At high temperature and small $\mu_B$, the symmetry is not broken and hence the equivalence works. 
This region is relevant for the heavy ion collision experiments. Note that this region, where the symmetry is intact, 
is exactly where the reweighting method works in principle (but of course difficult in practice). 
In such a region, one should not spend too much computational resource for the reweighting; 
just by ignoring the phase one can obtain reasonable results. 

%%%%%%%%%%%%%%%%%%%%%%%%%%%%%%%%%%%%%%%%%%%%%%%%%%%%%%%%%%%%%%%%%%%%
%%%%%%%%%%%%%%%%%%%%%%%%%%%%%%%%%%%%%%%%%%%%%%%%%%%%%%%%%%%%%%%%%%%%
%%%%%%%%%%%%%%%%%%%%%%%%%%%%%%%%%%%%%%%%%%%%%%%%%%%%%%%%%%%%%%%%%%%%
\section{Conclusion and outlook} \label{sec:conclusion}
\hspace{0.51cm}
%%%%%%%%%%%%%%%%%%%%%%%%%%%%%%%%%%%%%%%%%%%%%%%%%%%%%%%%%%%%%%%%%%%%
%%%%%%%%%%%%%%%%%%%%%%%%%%%%%%%%%%%%%%%%%%%%%%%%%%%%%%%%%%%%%%%%%%%%
%%%%%%%%%%%%%%%%%%%%%%%%%%%%%%%%%%%%%%%%%%%%%%%%%%%%%%%%%%%%%%%%%%%% 
 We have pointed out that QCD and  various QCD-like theories with chemical potentials are equivalent at large-$N_c$ 
 through the orbifold equivalence, at least to all order in perturbation theory. 
 QCD with the isospin chemical potential and $SO(2N_c)$/$Sp(2N_c)$ Yang-Mills with 
 the baryon chemical potential are equivalent everywhere in the $T$-$\mu$ plane, and furthermore, they are equivalent to 
 QCD with the baryon chemical potential outside the BEC-BCS crossover region.

Our result has immediate implication for the study of the chiral and deconfinement transitions in high-$T$, small-$\mu$ region. 
In this region it is reasonable to assume the $1/N_c$ correction is not very large (for example, 
as we have seen in \S~\ref{'tHooftVsVeneziano}, 
the leading corrections to the large-$N_c$ limit of the gluonic operators agree), 
and hence we can expect that 
the Monte-Carlo results of the QCD with isospin chemical potential describe the QCD with the baryon chemical potential 
with rather good accuracy. 
Furthermore, by using the $SO(2N_c)$ theory, one can study three-flavor theory without suffering from the sign problem.  
Similar study e.g. phase quenched simulation of $SU(3)$ 3-flavor QCD has been performed \cite{Kogut:2007mz}  
and the results suggest that  the QCD critical point does not exist. 
It is very important to study these sign-free theories numerically, further in detail,  
in order to find (or exclude) the QCD critical point\footnote{ 
Recently it has been argued that, in the strict large-$N_c$ limit, the QCD critical point cannot
exist outside the BEC/BCS crossover region of the phase-quenched theory \cite{Hidaka:2011jj}. 
Still it is important to study the theory numerically in order to see the details of the chiral transition, 
which provides us with a valuable information of the physics of the QCD with the baryon chemical potential 
which is hidden in the BEC/BCS crossover region of the phase-quenched theory.
}.

\bigskip
I would like to thank Aleksey Cherman, Carlos Hoyos, Andreas Karch, Daniel Robles-Llana,  Laurence Yaffe and Naoki Yamamoto 
for fruitful collaborations which this talk is based on. 
I also thank Brian Tiburzi for stimulating discussions and comments. 
My work is supported by Japan Society for the Promotion of Science Postdoctoral Fellowships for Research Abroad.


\begin{thebibliography}{99}

%%
%%  bibliographic items can be constructed using the LaTeX format in SPIRES:
%%    see    http://www.slac.stanford.edu/spires/hep/latex.html
%%  SPIRES will also supply the CITATION line information; please include it.
%%

%\cite{Cherman:2010jj}
\bibitem{Cherman:2010jj}
  A.~Cherman, M.~Hanada and D.~Robles-Llana,
  %``Orbifold equivalence and the sign problem at finite baryon density,''
  Phys.\ Rev.\ Lett.\  {\bf 106}, 091603 (2011)
  [arXiv:1009.1623 [hep-th]].
  %%CITATION = PRLTA,106,091603;%%


%\cite{Hanada:2011ju}
\bibitem{Hanada:2011ju}
  M.~Hanada and N.~Yamamoto,
 % ``Universality of Phases in QCD and QCD-like Theories,''
  arXiv:1103.5480 [hep-ph].
  %%CITATION = ARXIV:1103.5480;%%

%\cite{Hanada:2011ev}
\bibitem{Hanada:2011ev}
  M.~Hanada,
  %``Large-Nc equivalence and the sign problem at finite baryon density,''
  [arXiv:1109.6372 [hep-lat]].

%\cite{Kogut:2007mz}
\bibitem{Kogut:2007mz}
  J.~B.~Kogut and D.~K.~Sinclair,
  %``Lattice QCD at finite temperature and density in the phase-quenched
  %approximation,''
  Phys.\ Rev.\  D {\bf 77}, 114503 (2008)
  [arXiv:0712.2625 [hep-lat]].
  %%CITATION = PHRVA,D77,114503;%%

%\cite{deForcrand:2007uz}
\bibitem{deForcrand:2007uz}
  P.~de Forcrand, M.~A.~Stephanov and U.~Wenger,
  %``On the phase diagram of QCD at finite isospin density,''
  PoS {\bf LAT2007}, 237 (2007)
  [arXiv:0711.0023 [hep-lat]].
  %%CITATION = POSCI,LAT2007,237;%%


%\cite{Kachru:1998ys}
\bibitem{Kachru:1998ys}
  S.~Kachru and E.~Silverstein,
  %``4d conformal theories and strings on orbifolds,''
  Phys.\ Rev.\ Lett.\  {\bf 80}, 4855 (1998)
  [arXiv:hep-th/9802183].
  %%CITATION = PRLTA,80,4855;%%



%\cite{Bershadsky:1998mb}
\bibitem{Bershadsky:1998mb}
  M.~Bershadsky, Z.~Kakushadze and C.~Vafa,
  %``String expansion as large N expansion of gauge theories,''
  Nucl.\ Phys.\  B {\bf 523}, 59 (1998)
  [arXiv:hep-th/9803076].
  %%CITATION = NUPHA,B523,59;%%

  
%\cite{Bershadsky:1998cb}
\bibitem{Bershadsky:1998cb}
  M.~Bershadsky and A.~Johansen,
  %``Large N limit of orbifold field theories,''
  Nucl.\ Phys.\  B {\bf 536}, 141 (1998)
  [arXiv:hep-th/9803249].
  %%CITATION = NUPHA,B536,141;%%

  
%\cite{Kovtun:2003hr}
\bibitem{Kovtun:2003hr}
  P.~Kovtun, M.~Unsal and L.~G.~Yaffe,
  %``Non-perturbative equivalences among large N(c) gauge theories with  adjoint
  %and bifundamental matter fields,''
  JHEP {\bf 0312}, 034 (2003)
  [arXiv:hep-th/0311098].
  %%CITATION = JHEPA,0312,034;%%

%\cite{Kovtun:2004bz}
%\bibitem{Kovtun:2004bz}
  P.~Kovtun, M.~Unsal and L.~G.~Yaffe,
 % ``Necessary and sufficient conditions for nonperturbative equivalences of large N(c) orbifold gauge theories,''
  JHEP {\bf 0507}, 008 (2005)
  [arXiv:hep-th/0411177].
  %%CITATION = JHEPA,0507,008;%%



  
%\cite{Cherman:2011mh}
\bibitem{Cherman:2011mh}
  A.~Cherman and B.~C.~Tiburzi,
%  ``Orbifold equivalence for finite density QCD and effective field theory,''
  arXiv:1103.1639 [hep-th].
  %%CITATION = ARXIV:1103.1639;%%
  
  %\cite{Hidaka:2011jj}
\bibitem{Hidaka:2011jj}
  Y.~Hidaka, N.~Yamamoto,
  %``A No-Go Theorem for Critical Phenomena in Large-Nc QCD,''  
  [arXiv:1110.3044 [hep-ph]].


\end{thebibliography}
\end{document}